# GLOBAL LIFE PATTERNS:
# A PATTERN LANGUAGE TO DESIGN A PERSONAL GLOBAL LIFE

Ko Matsuzuka
Taichi Isaku
Satoshi Nishina
Takashi Iba

Keio University
5322 Endo, Fujisawa-shi, Kanagawa 252-0882 Japan
ko.matsuzuka@gmail.com

**ABSTRACT**

In this paper, we propose the Global Life Patterns, a methodology to support people to maintain individuality and design their own actions in respect to the "globalizing" society – a Global Life. Today, globalization requires each person to maintain individuality in respect to Globalization. Because methods to acquire such ability are tacit, a clarified methodology to live a Global Life is needed.This methodology is based on the format of a Pattern Language – specifically, this methodology is in the format of the Pattern Language 3.0(Iba, 2012).

**GLOBALIZATION**

Today, many would agree that Globalization is an intensifying phenomenon in society. The ripples caused by the bankruptcy of the Lehman Brothers spread beyond the financial market in the United States, and caused an international monetary crisis. Hopes as well as problems for a more global economy are currently being negotiated in the Trans-Pacific Partnership agreement, and multi-national debates over environmental challenges are heating up every day. Affairs which once were contained within individual countries are now inevitably initiating an endless chain of cause-and-effect reactions around the world.

So what is Globalization? The following sections will go through existing approaches of globalism by different persons in effort to grasp what it really means.

**Globalizing Society: a definition for Globalization**

Sociologist Roland Robertson mentions: "Globalization as a concept refers both to the compression of the world and the intensification of consciousness of the world as a whole." (Robertson, 1992) Sociologist Anthony Giddens also defines globalization: "the intensification of worldwide social relations which link distant localities in such a way that local happenings are shaped by events occurring many miles away and vice versa." (Giddens, 1991) Many would agree with both of these claims, especially after experiencing the financial crisis caused by the Lehman Brothers and the Greek government-debt crisis. In addition, the astounding progression and standardization of the Internet have connected individuals and organizations from around the globe with extremely low costs.

But what was behind "Globalization" that created it to become such a seemingly significant phenomenon? Thomas Friedman points out specific events in the past 30 years that intensified Globalization. After the fall of the Berlin wall, that allowed humanity to think of the world as a single space, Microsoft Windows 3.0 created a global computer interface, followed by the Netscape Revolution, allowing more people to connect with more other people from more different places in more different ways of collaboration. From then on, collaborative forms such as "outsourcing," "offshoring," "open-sourcing," "insourcing," and "supply-chaining" were possible, due to the availability of inexpensive, yet effective ways to access resources, such as fiber-optic cables and voice over Internet protocol. "It is this convergence – of new players, on a new playing field, developing new processes for horizontal collaboration – that I believe is the most important force shaping global economics and politics in the early 21$^{st}$ century." (Friedman, 2005)

However, note that Friedman mentions that individuals are globalizing, not so much because of an overall evolution in humanity's abilities, but because availability of resources which were not accessible before is rising rapidly.

This is why Friedman stresses the bitter fact that Americans and Western Europeans may lose their jobs to more "ambitious" personnel in other countries. If Globalization was only a change in humanity as a whole, costs of outsourcing would still be high, and even if ambitious personnel came to United States, Americans would know how to respond. However, because Globalization is the dynamism in the environment, rather than the individual, people become shocked by the fact that their field of battle has just gotten more diverse, and competitive.

This suggests that individuals *can* globalize rather than they *are* globalizing, because environments and its resources *are* globalizing. In other words, in modern society, it is rather the environment that is experiencing Globalization, more than humanity as a whole. Therefore, we propose that "Globalization" in current terms means the Globalization of society,

### The Complexity of the Global Society

So how did environments change, specifically? Arjun Appadurai claims that the new global cultural economy has to be seen as a complex, overlapping, disjunctive order that cannot any longer be understood in terms of existing center-periphery models, or migration theory, or traditional models of balance and trade. In Arjun Appadurai's terms, Friedman's type of environment can be comprehended as only a part of one of the 5 "landscapes" that characterizes the globalization of modern society – the "technoscape." The other "landscapes" include ethnoscapes, mediascapes, financescapes, and ideoscapes.

In addition, Appadurai explains that "These landscapes thus are the building blocks of what I would like to call imagined worlds, that is, the multiple worlds that are constituted by the historically situated imaginations of persons and groups spread around the world. An important fact of the world we live in today is that many persons on the globe live in such imagined worlds and thus are able to contest and sometimes even subvert the imagined worlds of the official mind and of the entrepreneurial mentality that surround them."(Appadurai, 2005) This suggests the intensification of complexity where philosophies are converging.

And I believe this convergence has reached its heights. Thomas Friedman divides the history of Globalization into 3 eras: Globalization 1.0 (1492 to 1800), Globalization 2.0 (1800 to 2000), and Globalization 3.0 (2000 to current). During Globalization 1.0, countries globalized for resources and imperial conquest. In Globalization 2.0, companies started globalizing for markets and labor. And in Globalization 3.0, there are enough resources for individuals and small groups to become globalized. The convergence of philosophies, or in Appadurai's terms, the "imagined worlds," can now happen at the individual level, rather than in more macro levels, such as communities or organizations.

What can be claimed from these insights presented in this chapter is that (a) we are currently experiencing a certain form of Globalization (b) Globalization can be defined as the Globalization of Society, more than the individual (c) the multiple aspects in Globalization are interrelated, meaning the existence of complexities in comprehending, predicting, and acting upon the Global Society.

## GLOBAL LIFE

### Emancipation

So what impact does Globalization have on the individual? The term "fluidity" is an important notion for Sociologist Zygmunt Bauman when explaining modern society. According to Bauman, in the past, ideologies that dictated society were solids, and the paradigm shifts that occurred were merely the melting of these solids in order to create new and improved solids to dictate society once again. However, according to Bauman, society has reached to a point where the melting and reconstruction of solids last longer than solids remaining as solids. *This is the fluidity of modern society, and it brings (or brought by) emancipation of individuals and communities.* "The liquidizing powers have moved from the 'system' to 'society', from politics' to 'life-policies' – or have descended from the 'macro' to the 'micro' level of cohabitation."(Bauman, 2000)

This theory seems to give hope to the individual, this may not be as glorious as it may seem. Emancipation forces the individual out of the orderly, predictable system that could have made his/her life much more simple. In an orderly, predictable system, the individual would know how to achieve success – moreover, understand what success is. This is because an orderly system offers a set of standards, a righteous path for those who are lost. Restaging, emancipation means that the individual is no longer under the protection of what his/her social system offers, and that he/she must take responsibility for the risks of his/her each and every choice.

### Individuality in the Global Society

However, while emancipation offers mixed blessings, our current world is indifferent. Bauman states: "What has been cut apart cannot be glued back together. Abandon all hope of totality, future as well as past, you who enter the world of fluid modernity. The time has arrived to announce, as Alain Touraine has recently done, 'the end of definition of the human being as a social being, defined by his or her place in society which determines his or her behavior and actions'."(Touraine, 1998) Modern society is becoming more and more fluid, meaning that emancipation is now less of a choice, but more of a bittersweet reality.

So what is one required to do in the fluidity, or liquidity, of the global modernity? Because members of the society are now no longer under the protection of social order brought upon by the environment, or the welfare state to negotiate the impact of risks on their lives such as a job loss, loss of pensions, they are forced into an era of individualization. In other words, each individual is required to cope with the fluidity of modern society, and acquire the ability to design his/her own personal life. As sociologist Richard Sennett notes in describing the generation gap in a family concerning the change in the concept of a career, "this conflict between family and work poses some questions about adult experience itself. How can long-term purposes be pursued in a short-term society? How can durable social relationships be sustained? … The conditions of the new economy feed instead on experience which drifts in time, from place to place, from job to job." (Sennett, 1998)

Restating, in society today, the individual must must be able to dismiss oneself from the order, predictability, titles for oneself and others, and the safety net their society used to guarantee, and brace one as an "individual" in its sincerest definition.

**The Concept of "Global Life"**
Given the significance of individualism in modern society, we would like to propose a way of life that maintains individuality in respect to the Global Society: a "Global Life".

Bauman states: "The 'leader' was a by-product, and a necessary supplement, of the world which aimed at the 'good society', or the 'right and proper' society however defined, and tried hard to hold its bad or improper alternatives at a distance. The 'liquid modern' world does neither." Therefore, in a society where each and every individual must decide his/her own choices, rather than following the footsteps or commands of a successful idol, it is crucial that the individuals do not imitate another's, but to refer to them, and design their own, personal actions or mentalities, given their specific environments. This is the first essence of a Global Life: diversity is a key factor, and what is important is the individual's ability to perceive their lives as their own, personal life, and that there is no master-plan. This means that no particular set of standards for a "successful" Global Life exist. Given the bitter side of emancipation mentioned in Section 3.1, the individual must decide what "success" is to him/her.

Though the term "Global" may not be completely definable, the world is undoubtedly in need of a way to cope with the term. As shown in the previous section, there are already many interpretations of the word. The existence of the multiple interpretations itself suggests that the word simply has not a single correct definition but holds many aspects which each can be understood differently.

With such many sides to its concept, the word then holds the danger of causing misunderstandings between people. Furthermore, since the term covers such a wide variety of concepts, its use is starting to become somewhat intuitive. When reconsidered, it is very likely that many of us are misunderstanding the concept. Although sociologists have found aspects of the global age, the concept consists of infinite more aspects, which are yet to be found. This does not suggest that we should stop the exploration for its definition, but rather hints the very fact that a unique Global Life exists for each individual on the planet.

Thus it is significant for each individual to design his/her own Global Life, because it eventually leads to the integration of more aspects into the global age. Therefore, every individual, as members of the Earth, can share their own perceptions they have designed or discovered, in order to create a more diverse, and creative global society.

So how does one integrate the concept of "global" into his/her daily actions or mentalities, in order to strive in the global environment? In a state where the individual must act accordingly to his/her lifestyle and philosophy, it is necessary for the individual to comprehend his/her current situation, discover a problem which emerge within that situation, and find a solution to solve the problem. This is the second essence in a Global Life. This context-problem-solution approach is based on the "Alexandrian form" of a methodology called Pattern Language, first presented by architect: Christopher Alexander. Alexander claimed that the contexts in designing architecture were highly complex, and proposed the necessity of a problem-finding-and-problem-solving structure to create architecture based on solely for inhabitants' use. Further description of Pattern Language is mentioned in the following chapters.

The significance here is that the flow of the context-problem-solution structure is determined by the problem finding and problem solving of the individual. In the complexity of the Global Society, and when forced into emancipation, the individual must decide what "success" is for himself/herself, and how far he/she stands in terms of that success. This is the act of comprehending the context, and is the first step in reaching the personal success. The difference between where the individual stands in terms of his/her success suggests that there are still problems to be solved. This is when the art of problem-finding comes in, in order to fill the gap between the context and its potential problems. The individual must comprehend what aspects of the complexity of Globalization are in the current, specific context, and what kinds of mechanisms, structured by the aspects, have the potential to cause a problem. Then, in order to create, and execute a solution for the problem, the individual now must go through a process of problem-solving.

In other words, a "Global Life" is defined as a life where the individual (a) constantly rediscovers what "success" is in his/her very own, personal life. This means becoming independent from the restrictions *or* privileges that society or community(s) may have provided before (b) discerns his/her context, finds the problem, and solves it with his/her own version of a solution, in order to achieve his/her own version of "success."

However, aspects of the current global society are mainly derived from individuals who are already living a Global Life. In modern society, there are arguably those who have their own perception of the concept of "Global," and even live their own version of a Global Life. Some of them, such as the sociologists referred to in previous sections of this paper, manage to open their perception to the public. And some of them even demonstrate a Global Life, such as members of the World Economic Forum in Davos. These perceptions create the concept of "Global" today, and allowed us to understand the fluidity, emancipation, individuality, etc. of modern society in a much deeper comprehension.

However, although these perceptions of these personnel offer an insight to what the Global Society is, and the significance of it, the concept of "Global" in their terms is only the tip of an iceberg. In current society, these perceptions created by sociologists, or members of the World Economic Forum, and their lives are glorified as the pinnacle of a Global Life, in which creates a misconception that their specific actions are parts of a master-plan to design a Global Life. However, as mentioned in describing the first essence of a Global Life, this "knowledge to live a Global Life" is often times tacit, and even if they are open to the public, imitation may not lead to another's Global Life. As a result, individuals are lost on attempt to design their own Global Life, because in the era of emancipation and individuality, they can refer to, but cannot imitate one another.

Thus it is crucial to create a methodology that does not have a master plan, and provokes the individual to design and discover his/her personal Global Life.

## A PROPOSAL OF GLOBAL LIFE PATTERNS

### The Format of Pattern Language

To solve this problem, we propose a pattern language as a methodology to a help individuals design their Global Life.

The original idea of pattern languages was proposed by an architect Christopher Alexander. Alexander was critical for an expert to design according to his/her personal conception of beauty, rather than the community's conceptions. This methodology was written in 253 'patterns' which shows tips and their beauty in building architecture, designing an office, or a workshop, or a public building, or even neighborhoods with others, in a manner that is comprehensible and accessible to non-architects. Each pattern was created under the same "Alexandrian form" mentioned in the previous section.

At the root of these patterns was the idea that they were written as a language. Alexander stated that in designing their environments people always rely on certain "languages," which, like the languages we speak, allow them to articulate and communicate an infinite variety of designs within a forma system which gives them coherence. "Patterns," the units of this language, are answers to design problems in architecture, and their inter-relational flux creates a certain quality, like in literature.

There are two significances of this Pattern Language proposed by Alexander, in terms of designing a Global Life. First, it empowers the individual to decide what they really need, or look forward to, rather than idolizing a set of fixed standards, or a successful example. Its idea that wonderful places of the world were not made by architects but by the people, and its "language" format reflects the essences of designing a Global Life. Second, its format of writing the problem statement, discussion, and solution give the user insight on what to expect and how to solve potential problems. In addition, the Title and its overall writing style gives even the inexperienced individuals effective and accessible insights to designing seemingly professional work.

This structure of a bridge between experts and users opened what was once tacit in the experts' minds. Ten years after the creation of the pattern language for architecture, the idea of pattern languages was adopted in the field of software design.

And, recently, the fields where pattern languages are applied are expanding gradually. A format relevant to creating a pattern language for designing a Global Life is called Pattern Language 3.0, presented by Takashi Iba (Iba, 2012). Iba identifies the generation of Pattern Language when created for architecture Pattern Language 1.0, the generation when created for software design Pattern Language 2.0.

Iba states that one of the most crucial shifts from previous forms to Pattern Language 3.0 was that it designed human actions. While Pattern Language 1.0 and 2.0 supported designing what was separate from the user, Pattern Language 3.0 supported designing an object inseparable from the user. In addition, as he describes the act of design amongst the three generations, "The new 3.0 stage shows a new aspect. Pattern languages in this stage can be said that it is constantly being designed. Unlike architecture that has a concrete border that marks before and after the designing process, or software design where each version of the codes can be marked with the release of an update, human actions are something that is put into practice both constantly and continuously." In addition, while Pattern Language 1.0 served as a "lingua franca" between designers and users, and Pattern Language 2.0 was between experts and non-experts, Pattern Language 3.0 connects all kinds of people with all kinds of different experiences. 3.0 patterns support communication between individuals so they can understand a specific aspect of "global." Because a Global Life derives from designing human behavior, it is radical to write it in the Pattern Language 3.0 format.

### **Function of Global Life Patterns**

We propose the Global Life Patterns, a pattern language to support people design and live a Global Life. A pattern language is a set of "patterns" which each scribe out the complex relationships of a person's knowledge, especially of those which are tacit and usually embedded deeply into the person's mind and actions. Through a mining process, in this case through interviews, these knowledge are verbalized and scribed out. These kinds of knowledge often come in the form of a solution to a complex problem. Thus, these patterns are written in a rather strict format, containing the "context" in which a "problem" occurs, and the "solution" to this problem is the knowledge which has value to be written out. This set of information is grouped together as a "pattern", and then is given a name. The complex knowledge then can be referred to by the pattern name, making communication about the idea to occur easily. Having the patterns in their mind would also allow users to cut out and recognize patterns out of otherwise unnoticed sequence of events. There are multiple patterns in a pattern language, and the users would choose and combine patterns out of the language to design their own actions.

Each pattern in the Global Life Patterns is mined out through interviews with people who already are able to live a Global Life. These patterns are meant to give its users a new perspective to look at the concept of "global". By having the patterns, people would be able to inherit the knowledge of the experts and use them as an abstract guideline for designing their own Global Life according to their contexts and philosophies.

Once their own version of a Global Life is formed, this new Global Life will have a created a new aspect of the "global" concept. This person has now become a part of the concept, and will add on a new aspect to its concept. Once multiple users discover and act upon this mechanism, the Global Lives of all the individuals will be added onto the "global" concept. These are views which were not reflected onto the larger concept before. When this is achieved, the concept of the word: "global" will more accurately reflect the visions of all the individuals on the planet, and less will be left behind from the "global" concept.

This designing the "global" concept and the individual Global Lives is a rather dynamic process. By the views of the individuals becoming added onto the concept of the "global," the concept itself will grow dynamically with the process. From these new versions of the concept, new patterns can additionally be written. The new patterns would then give the users more, and ideally available, perspectives, and they would be able to redesign their Global Life from the discovering the other concepts. Consequently, both the concept of "global" and the individual Global Lives grows through the process, through influencing each other.

### **COLLABORATIVE PROCESS OF CREATING GLOBAL LIFE PATTERNS**

In order to integrate a variety of perspectives, we will create the Global Life Patterns through a collaborative process. In addition, we plan to build a mechanism to make this collaboration open to the public, with a Collaborative Innovation Network. Nurturing a Collaborative Innovation Network for the Global Life Patterns is significant, because its concept emphasizes the diversity of dynamic perspectives.

**Creation Process of the Global Life Patterns**

The creation process can be broken down into 3 main phases: pattern mining, visual clustering, and pattern writing. Although these phases may have a sequential impression, they are rather integrated in the process spontaneously and repetitively.

In pattern mining, the creators explore experiences from a variety of individuals, and discern any tacit knowledge within those experiences, through communication. Through this exploration, creators identify and extract specific actions or mindsets which seem to be relevant in the individual's Global Life. These specific actions or mindsets, called the "seeds of patterns," have the potential to be written as patterns.

In visual clustering, the creators brainstorm the seeds of patterns derived from the pattern mining process, and cluster together those which are similar or relevant. In this process, the creators consider every derived seed of patterns simultaneously. In order to make this possible, the creators utilize stationaries such as sticky notes in order to write the seeds of patterns down for brainstorming, and physically move and place them around on a blank poster paper.

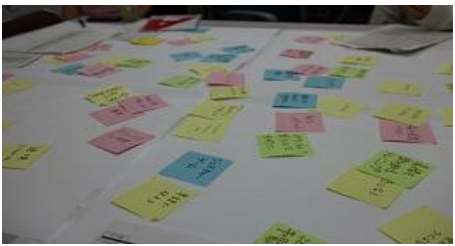

*Figure 2: Brainstorming and clustering actions from interviewees.*

In the pattern writing process, the clusters created from the visual clustering process are composed into single pattern formats. This means that each cluster is broken down into the context-problem-solution format of pattern languages. To be specific, each pattern contains a pattern name, a single sentence explanation, an illustration, a case, context, problem, forces, solution, actions, a rationale, and a consequence. The pattern name, single sentence, and illustration allows the user to recall the pattern easily and utilize the pattern as a common language through communication. The context, problem and solution are as explained in previous sections. The forces explain why such problems occur in the given context. The actions demonstrate detailed ways to execute the solution. The rationale presents the theoretical principles which explain why such solution and actions solve the problems. The consequence presents possible outcomes when the pattern is used.

The team of creators of the Global Life Patterns is formed by members of the Takashi Iba laboratory (including the authors of this paper). The members create the patterns through the three main phases, and weekly, meetings are held to discuss the concept of the Global Life Patterns and to brainstorm prototypes for patterns.

Members also interview various individuals to find out what perspectives they have on a global life. So far, we have interviewed eleven Japanese individuals, who interact with foreign culture or people on a regular basis. These interviewees include: a chairman of a non-profit organization, professor in the field of linguistic semantics, a professor in the field of global internet governance(shown in Figure 4.3), a professor in the field of global education, an employee of a trading company, an entrepreneur in the internet finance business, a member and organizer of the World Economic Forum. Each interview is around an hour and a half long. Common questions are "What do you think the concept of "global" is?" "What actions do you take in the Globalizing society?" "Is there any specific persons you feel that they are 'Global'?" However, the questions always vary depending on the interviewee's background, or the emerging context during the interviews. For example, we discussed more on the semantics of terms such as "global," "international," "life" to the professor in the field of linguistic semantics, while we discussed more concerning specific business situations to employees in the global departments of companies.

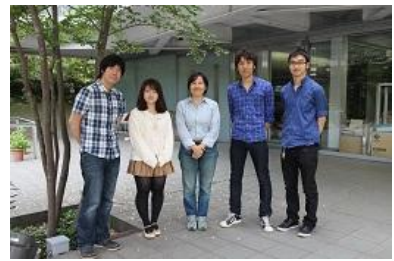

*Figure 3: Interviewee (middle), a professor of Digital Communication and Education Environment at Keio University.*

We record every interview entirely, so we can write the entire conversation down, and to be able to recall every single detail of the interview. From this dialogue, the creators extract specific actions or mindsets (or seeds of patterns) that we feel is supporting the interviewee to design his/her Global Life.

**Utilizing the Collaborative Innovation Network**

These interviewees are also members of the COIN, along with other members of the Takashi Iba Laboratory. These members avidly contribute to the creation of Global Life Patterns through participating in discussions in the multiple interviews and meetings held every week. Through these discussions, members give their own individual knowledge derived from experiences and expertise. The creators integrate the COIN's contributions into creating the concept of "global", and prototypes of patterns. For example, there were multiple instances during interviews where we realized the existence of completely different perspectives of the concept of "global." In other instances, we presented our ideas on a Global Life, or answered questions *from* the interviewees. By designing interviews so that the interviewee can think about the concept of global with us, the interviews itself became creative and new perspectives were discovered.

**CONCLUSION**

In a world where globalization is intensifying, and individuality becomes crucial, the Global Life Patterns will serve as a highly effective tool for communication to support an individual to design his/her own Global Life. This is primarily because it provokes an individual to discover and design his/her own way of a Global Life. This will provoke individuals can use these patterns as a medium to communicate, and a tool to design and maintain individuality in the Global Society. As a result, individuals designing their own Global Life will ultimately lead to a Global Society that comprehends the variety of perspectives of "Globalization."

In the future, we plan to create more patterns, and also to rewrite existing patterns. In order to integrate more perspectives, we will continue interviewing – especially to non-Japanese individuals. We will also contact individuals interviewed before, and present them the prototypes of patterns. This will allow them to realize their specific actions or mentalities as patterns for a Global Life. We will also present these patterns in workshop where participants can physically share their experiences or specific mentalities in their Global Life.

**APPENDIX**

Below is a pattern from the Global Life Patterns.

### Friend's Friend

Going just one step further to your acquaintance's acquaintance can expand your world.

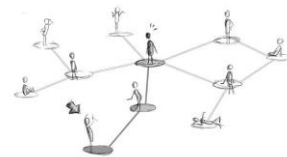

**Story**

An author of the Global Life Patterns was struggling to find an environment where he could work at a worldwide scale. Even if he looked around his acquaintances, he could not see building an environment at a worldwide scale with any of the relationships, and even if he did, he only saw relationships that required too much money or time. He started asking many of his acquaintances from his research group and club. One day, his friend told him how he was taking a course where employees of a company in the Internet business, based in the United States, was teaching. So the author attended the next class with his friend, and approached an employee together. Through his friend, the author was able to introduce himself, and express his passion. As a result, he was accepted as an internship, where he was offered a job to send information about the Japanese branch worldwide, and even a business trip abroad. Through his acquaintance's acquaintance, he was able to learn how to respond to foreign clients, and understand height of the quality expected in the worldwide level.

**Context**
**In order to prepare for having to work in an environment you are not used to, such as negotiating with a foreign client, you are finding an environment to work where the country, language, or culture is different than usual.**

▼ In this context

**Problem**
**Because you cannot find an environment you want, you give-up, thinking that you cannot stretch your possibilities.**
When you are finding a person who can give you an environment to work, you would probably ask your acquaintances first. However, the amount of acquaintances an individual can have is limited. Moreover, the possibility of knowing someone who can build a better environment to work with you, or introduce such environment is usually low. This is because of the fact that humans tend to gather around those who are similar to each other, and because such communities lack in diversity, it is difficult for each individual to receive opportunities that lead them farther than their current state. In addition, trying to find an environment or opportunity alone is difficult, because it is rather indefinite.

▼ Therefore

**Solution**
**Even if your friends and acquaintances cannot introduce the environment you desire, take one step further to their friends and acquaintances.**
Even if you are not connected with friends and acquaintances who can introduce the environment you desire, open your eyes to who they are connected to. For example, ask an acquaintance or friend that has a high possibility of knowing someone from the environment you desire. Your seniors in school or your boss in your company may introduce their acquaintances, knowing what kind of environment you seek. Or, ask someone you are acquainted with, who has many friends and acquaintances. Individuals who are in various communities are likely to be acquainted with people from a diversity of backgrounds. If you utilize social networking systems, you can search for acquaintances and their connections inexpensively. Like so, take one step further to your acquaintance's acquaintances.

▼Because

**Rationale**
In modern society, a social theory called: "Six degrees of Separation" exists, stating that everyone, on average, is six or fewer steps away, by way of introduction, from any other person in the world, so that a chain of "a friend of a friend" statements can be made to connect any two people in a maximum of six steps. The existence of individuals who have a significant amount of acquaintances, called a "hub," is one of the reasons this phenomenon is possible.
Lets take an ordinary university student in Japan, for example. The student can probably connect to a professor directly, or at least one acquaintance. That professor is probably acquainted with other professors or the Dean of the university, who are connected with researchers, professors, or officers of companies and organizations around the world. Although the university student may not have such acquaintances around him, he was able to connect to activists around the world through only two to three acquaintances. In addition, asking the professor at the university he attended was rather inexpensive.
Many say that people are getting even closer than six acquaintances, due to the development of social networking systems. By utilizing such principle, and opening your eyes to your acquaintance's acquaintances, your chances of finding the person you desire will significantly rise, even if you may not have anyone who can introduce you to the environment you desire.

▼Consequently

**Consequence**
Your range of finding people significantly expands, and your chances of finding a new environment increases. In addition, because you both have the same acquaintance, your acquaintance's acquaintance is likely to have some sort of relationship with you, and has the possibility that your jobs or backgrounds are actually related. Thus, you can do the new things you wanted to without having to completely leave your current environment behind.